# Guided plasmon modes of triangular and inverted triangular cross section silver nanoridges

Zeyu Pan,[1] Junpeng Guo,[1,*] Richard Soref,[2] Walter Buchwald,[3] and Greg Sun[2]

[1]Department of Electrical and Computer Engineering, University of Alabama in Huntsville, Huntsville, Alabama 35899, USA

[2]Department of Physics, University of Massachusetts at Boston, Boston, Massachusetts 02125, USA

[3] Solid State Scientific Corporation, Hollis, New Hampshire 03049, USA

*Corresponding author: guoj@uah.edu

Propagating two-dimensional plasmon modes guided along silver nanoridge waveguides with triangular and inverted triangular cross sections are investigated in this paper. Mode field profiles, dispersion curves, propagation distances, and figure-of-merits of the plasmon ridge modes are calculated for silver nanoridge waveguides with various triangular and inverted triangular waveguide cross sections. It is found that the triangular cross section nanoridge waveguide, if designed properly, can have longer propagation distance and higher figure-of-merit than the flat-top nanoridge waveguide of the same width. When the triangle height of the nanoridge is high, the mode approaches to the small angle wedge mode. An inverted triangular cross section nanoridge mode can be considered as a hybrid mode of two metal wedge plasmon modes. When inverted triangle depth increases, the propagation distance and the figure-of-merit decrease dramatically, suggesting the poorer performance when compared to the flat-top nanoridge plasmon waveguide.

## 1. Introduction

Surface plasmons are free electron density oscillations on the metal-dielectric interfaces, and can propagate along the metal-dielectric boundaries in the form of surface plasmon waves with tightly confined sub-wavelength modes [1-3]. Surface plasmon modes in various waveguide structures, such as thin metal films [4-8], finite width thin film metal stripes and metal wires [9-17], trenches in metal surface [18-29], metal dielectric layer structures [30-39], dielectric-loaded metal films [40-46] and metal wedges [25, 26, 47-52] have been extensively investigated. However, these waveguides that support tightly confined modes also suffer from high attenuation, rendering only limited applications. The quest for low loss but high confinement waveguides has inspired much research effort in the plasmonics community of various shapes and dimensions. Recently, a gold nanoridge was fabricated using the focused ion beam milling method [52] which confirms that metal nanoridges can support surface plasmon propagating modes.

The mode properties of flat-top silver metal nanoridge plasmon waveguide have been investigated previously [53]. In this paper, comprehensive numerical investigations on the plasmon modes supported by triangular and inverted triangular cross section silver nanoridge waveguides are presented. The motivation for investigating such metal nanoridge waveguides is to find an optimal structure that gives longer propagation distance and higher figure-of-merit. We calculate the mode field distributions, mode indices, dispersion curves, propagation distances, mode sizes, and the figure-of-merits of triangular and inverted triangular silver nanoridges with different heights and depths as well as their dependence on the wavelength.

## 2. Triangular cross section silver nanoridge plasmon waveguides

In this section, we systematically investigate the triangular silver nanoridge waveguides. The 3D view and cross section of such a triangular silver nanoridge are shown in Fig. 1(a) and Fig. 1(b), respectively, where the nano-scale metal ridge is extended in the *z*-direction with its width (*w*) in the *x*-direction and height (*h*) in the *y*-direction. The plasmon wave propagates along the ridge top in the *z*-direction. We assume the height of ridge is sufficiently large so that the substrate does not influence the ridge plasmon mode. Our previous work on the plasmon modes supported by the flat-top silver nanoridges [53] has found that the ridge width of 120 *nm* is optimal which is a compromise between the propagation distance and the mode confinement. In this study, we thus fix the ridge width at 120 *nm*, for both triangular and inverted triangular nanoridge waveguides.

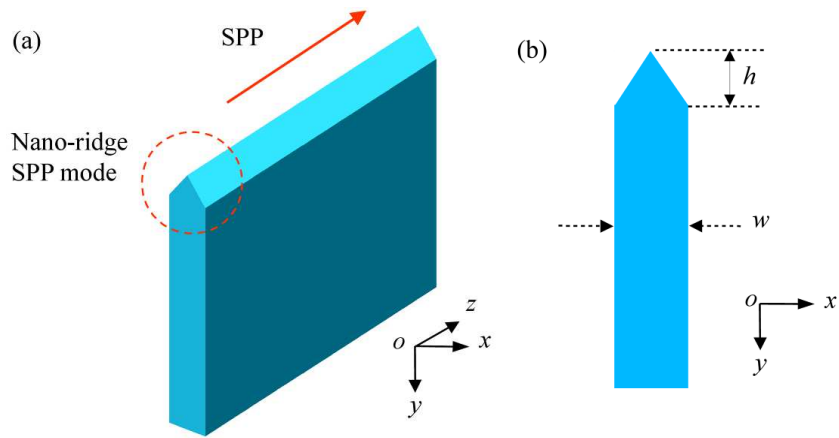

Fig. 1. (Color online) (a) 3D view of the triangular nanoridge plasmon waveguide. (b) Cross section of the triangular nanoridge plasmon waveguide.

We consider the situation where the surrounding medium of the ridge waveguides is air ($\varepsilon_d$ = 1.0), and the ridges are made of silver with its electric permittivity $\varepsilon_m$=-127.5-5.3$j$ at the telecommunication wavelength of 1.55 $\mu m$ [54]. For the flat-top ridge with width of 120 $nm$, the mode index is equal to $n_{eff}$=1.018-0.00074$j$, and the attenuation coefficient is 261.87 $dB/cm$. While this study investigates silver as the ridge material, the analysis can obviously be extended to other types of high electron density materials such as heavily doped semiconductors [55, 56] in different wavelength ranges.

We calculated the electric field intensity distributions of the fundamental plasmon propagating modes supported by triangular nanoridges of different triangle heights at the wavelength of 1.55 $\mu m$. The widths of the nanoridges are all at 120 $nm$. Fig. 2(a) is the electric field intensity distribution of the flat-top nanoridge plasmon waveguide mode. The mode of the flat-top nanoridge has two hot spots located at each corner of the ridge, which can be considered as a hybrid mode of two 90° wedge plasmon modes [53]. Fig. 2(b) is the electric field intensity distribution of the nanoridge plasmon mode of the nanoridge with the triangle height of 20 $nm$. Fig. 2(c) is the electric field intensity distribution of the nanoridge plasmon mode of the nanoridge with the triangle height of 34.6 $nm$. When the triangle height is equal to 34.6 $nm$, the three wedge angles of the triangular cross section nanoridge are the same at 120 degree. Fig. 2(d)-(f) are the electric field mode intensity profiles of the nanoridge plasmon waveguides with the triangle height of 40 $nm$, 60 $nm$, and 80 $nm$, respectively. It can be seen that for a triangular nanoridge with height between 20 $nm$ and 50 $nm$, the mode spans the entire ridge and is composed of three hot spots located at three corners of the ridge, which can be considered as three coupled wedge plasmon modes. The mode progressively converges towards the tip of the triangle with the tighter mode confinement as the triangle height increases. As the height increases to above 60 *nm*, the mode converges to only one hot spot at the tip, which can be considered as a small angle wedge mode. We also find that the mode energies are distributed over three wedges with small triangle height and the modes are not very sensitive to the change of the triangle height. But once the plasmon mode is concentrated at the tip of the triangle ridge when the height is greater than 60 *nm*, the hot spot at the top wedge rapidly intensifies with the increase of the triangle height.

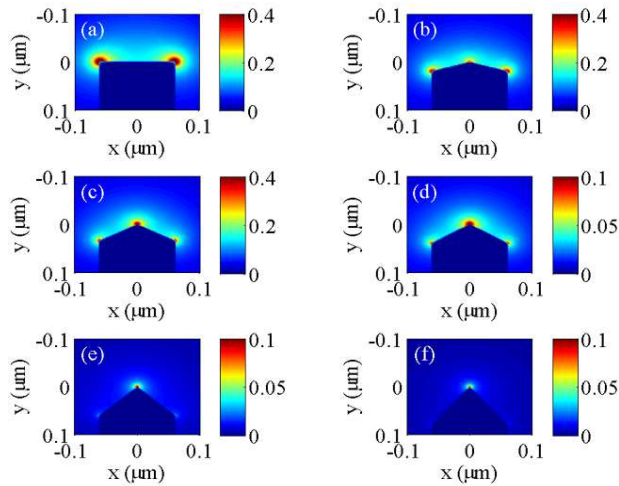

Fig. 2. (Color online) Electric field intensity distributions of the 120 *nm* wide triangular nanoridge plasmon waveguide with triangle height equal to (a) 0 *nm*, (b) 20 *nm*, (c) 34.6 *nm*, (d) 40 *nm*, (e) 60 *nm*, and (f) 80 *nm* at 1.55 $\mu m$ wavelength.

The mode dispersion curves for triangular silver nanoridges of different triangle heights are shown in Fig. 3. The black solid line and dashed line are the dispersion curves of the light line in air, and the silver-air flat plasmon mode, respectively. The triangle height increases from the zero (the flat-top ridge) to 120 *nm*. As the frequency increases, the plasmon mode dispersion curves of all ridges drift away from the light line in air, leading to slower group velocity and tighter mode confinement. This effect is far more pronounced for ridges with larger height. When the triangular height increases from the zero, the dispersion curve moves toward the light line first, that indicates the reduction of the mode confinement and propagation attenuation. Further increase of the triangle height moves the dispersion curve away from the light line, which indicates the increase of the mode confinement and propagation attenuation. This trend can be clearly seen from the insert in the Fig. 3. The red line is the dispersion curve of the flat-top nanoridge waveguide. The blue line is the dispersion curve of the nanoridge waveguide with 30 *nm* triangle height. The green line is the dispersion curve of the nanoridge waveguide with 60 *nm* triangle height. The magenta and cyan dashed line is the dispersion curve of the nanoridge waveguide with 90 *nm* and 120 *nm* triangle height respectively. Dispersion curves of ridges with height between 20 – 50 *nm* are not very sensitive to triangle height change. Beyond a height of 60 *nm*, the mode intensity is more concentrated to the tip of the triangular ridge. The dispersion curve moves rapidly away from the light line when the height of the triangle ridge continues to increase. The ridge plasmon mode gradually approaches to a wedge plasmon mode.

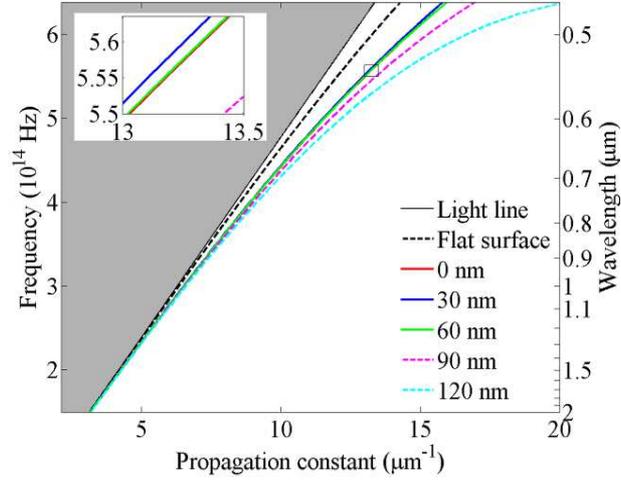

Fig. 3. (Color online) Dispersion curves of the silver triangular nanoridge plasmon waveguides of different triangle heights and the comparison with the dispersion curve of the silver-air flat plasmon mode.

The real and imaginary part of the triangular nanoridge mode index versus the wavelength for several different triangle heights ($h = 0$ *nm*, 30 *nm*, 60 *nm*, 90 *nm*, and 120 *nm*) are plotted in Fig. 4, where the black dashed lines are those of the silver-air flat plasmon mode. As the wavelength increases, both the real and imaginary part of the triangular nanoridge mode index are reduced, indicating the reduction of the mode confinement and propagation loss. We can also see when the triangle height increases from the zero, both of the real and imaginary part of the triangular nanoridge mode index first reduce, and further increase of the triangle height increases those of the triangular nanoridge mode index, which is consistent with results in Figs. 2 and 3.

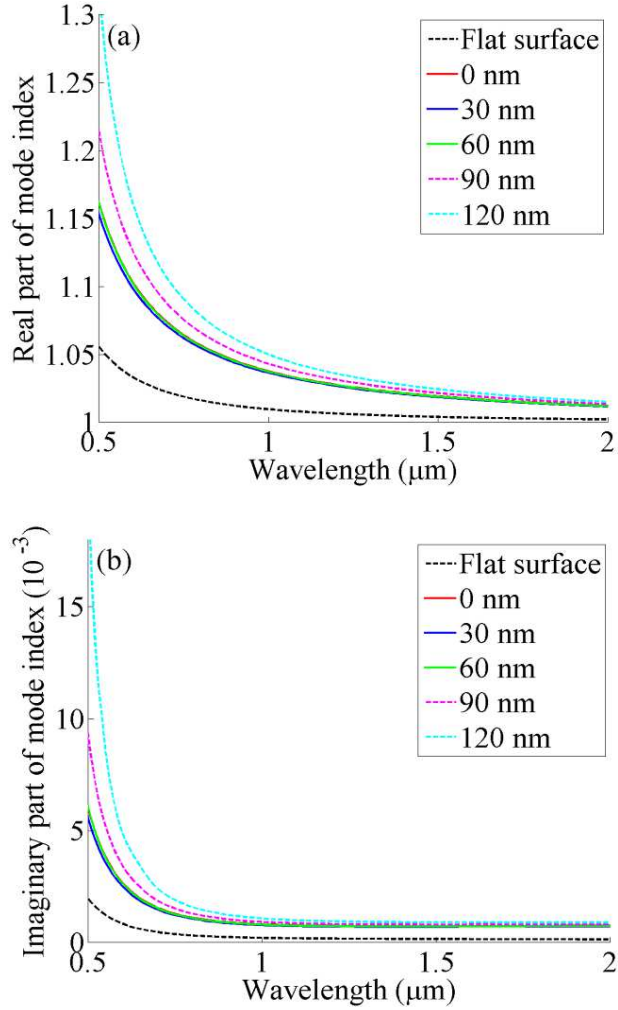

Fig. 4. (Color online) (a) Real part of the mode index versus the wavelength for different triangle heights and the comparison with the real part of the silver-air flat plasmon mode index. (b) Imaginary part of the mode index versus the wavelength for different triangle heights and the comparison with the imaginary part of the silver-air flat plasmon mode index.

The real part and imaginary part of the mode index versus the triangle height at 1.55 $\mu m$ wavelength are shown in Fig. 5. Interestingly, both the real and the imaginary part of the triangular nanoridge mode index reduce with the initial increase of the triangle height, indicating a reduction in attenuation and confinement. The rapid increase of mode index thereafter suggests, however, quick rise of attenuation and confinement. This suggests a minimum value for both mode confinement and propagation loss at a triangle height of roughly 40 *nm*, beyond which both the real and imaginary part of the mode index increase dramatically.

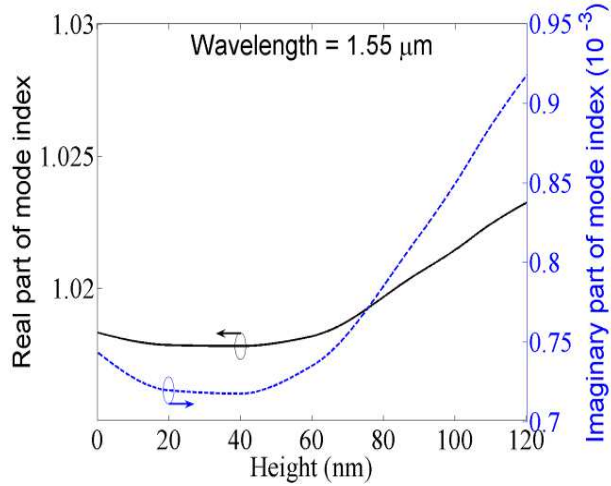

Fig. 5. (Color online) Real and imaginary part of the mode index as a function of the triangle height at 1.55 *μm* wavelength.

In general, the propagation surface plasmon modes can be characterized by a complex wave vector $\beta_z = n_{eff} k_0 = \beta - j\alpha$, along the z-direction, where $\beta$ is the phase propagation constant of the mode, and $\alpha$ is the attenuation constant. The propagation distance is defined as the distance where the mode intensity attenuates to 1/*e* of its initial value, i.e. $L_p = 1/(2\alpha)$. The propagation distances are calculated for triangular nanoridge plasmon waveguides of different triangle heights at different free space wavelength. The results are shown in Fig. 6 (a) and (b). It can be seen that as the triangle height increases, the propagation distance increases first, and then reaches a plateau for the height between 20 - 50 *nm*. After that, the propagation distance decreases rapidly. As the wavelength increases, the propagation distance also increases. As discussed in the Ref. [26], for the wedge plamson mode, the loss increases dramitically as the wedge angle decreases. When the triangle height is between 20 – 50 *nm*, the mode energy is more distributed among three wedges, and the nanoridge plasmon mode can be considered as a coupled wedge mode. Between that triangle height range, the coupled wedge plasmon modes of the triangular nanoridge have smaller loss, so the nanoridge plasmon mode has the longest propagation distance.

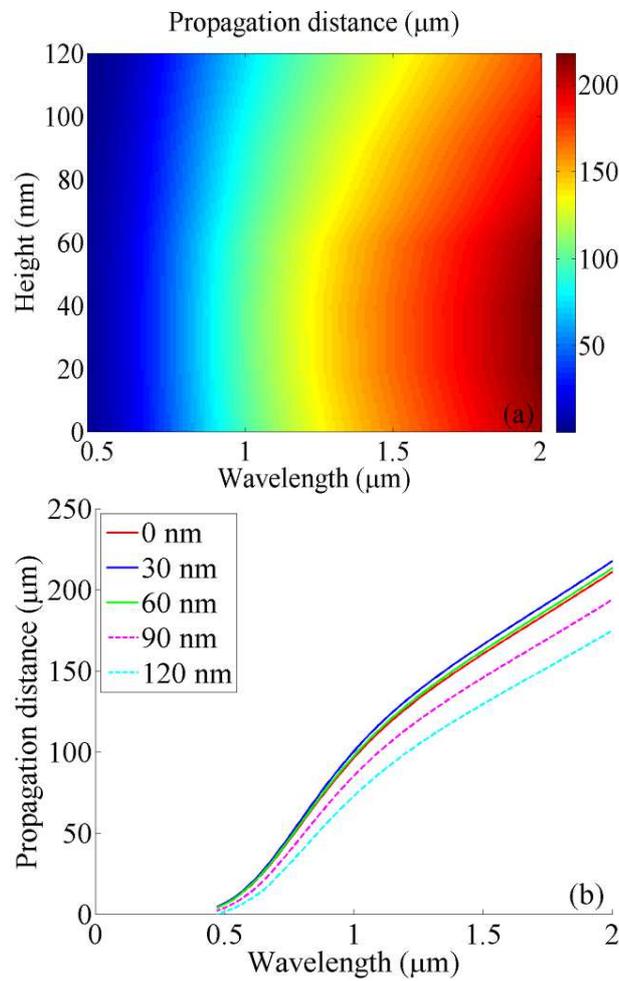

Fig. 6. (Color online) (a) Propagation distance versus the wavelength and triangle height. (b) Propagation distance of the triangular nanoridge mode as a function of free space wavelength for different triangle heights.

Outside the metal in the surrounding dielectric medium, the transverse component of the complex wave vector is given by $\beta_\perp = \gamma - j\delta$, where $\gamma$ and $\delta$ describe the oscillation and the field decay in the transverse direction respectively. It follows from Maxwell's equations that the complex propagation constants in the propagation direction and the transverse directions are related as:

$$(\beta - j\alpha)^2 + (\gamma - j\delta)^2 = \varepsilon_d k_o^2 \qquad (1)$$

where $k_0$ is the mode propagation constant in the free space, and $\varepsilon_d$ is the dielectric constant of the surrounding dielectric. Solving Eq. 1, we can obtain $\gamma$ and $\delta$, which can be used to define the mode size as $1/(2\delta)+w+1/(2\delta)=w+1/\delta$. The mode size of the triangular nanoridge waveguide versus the free space wavelength and triangle height is shown in Fig. 7(a), while Fig. 7(b) shows the mode size versus the free space wavelength for triangular nanoridge of different triangle heights. Here is seen that the mode size increases, as the wavelength increases. Once again, for the triangle height smaller than 20 *nm*, the mode size slightly increases, but when the height increases higher than 60 *nm*, the mode size decreases rapidly.

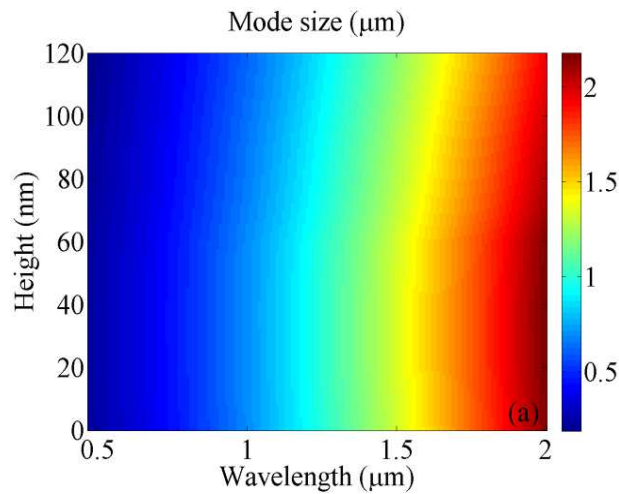

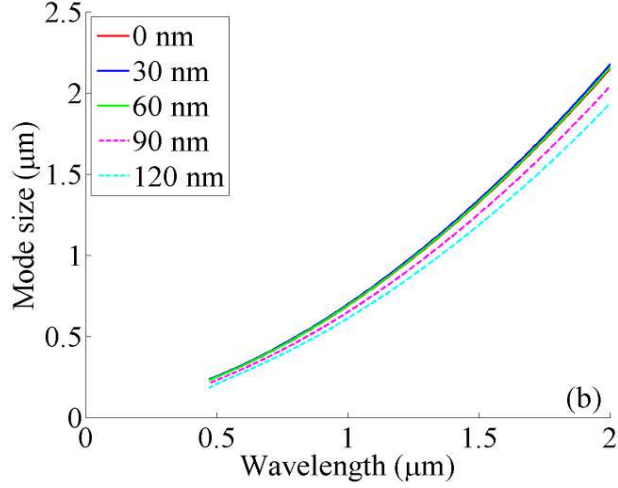

Fig. 7. (Color online) (a) Mode size versus the wavelength and triangle height. (b) Mode size of the triangular nanoridge plasmon mode as a function of free space wavelength for different triangle heights.

While it is always desirable to obtain surface plasmon waveguides with tight confinement and low attenuation, the reality is that there is always a trade-off between the propagation attenuation and the mode confinement [12, 57]. While the small mode confinement is the merit, the attenuation is the cost. Figure-of-merits were proposed to characterize the trade-off between the attenuation and the confinement [58, 59]. Here, we define the figure-of-merit of the nanoridge plasmon waveguide as the ratio of the propagation distance over the mode size:

$$FoM = (1/2\alpha)/(w + 1/\delta) \qquad (2)$$

We calculated the figure-of-merit of the triangular nanoridge waveguide versus the wavelength and the triangle height. The results are shown in Fig. 8(a). Fig. 8(b) shows the figure-of-merit versus the wavelength for several different triangle heights. It can be seen from Fig. 8(a-b) that the figure-of-merit reaches the maximum at 1.05 $\mu m$ wavelength for all triangular nanoridge waveguides of various heights considered in this study. It also can be seen that higher figure-of-merits are achieved for ridges with height between 20 – 50 *nm*.

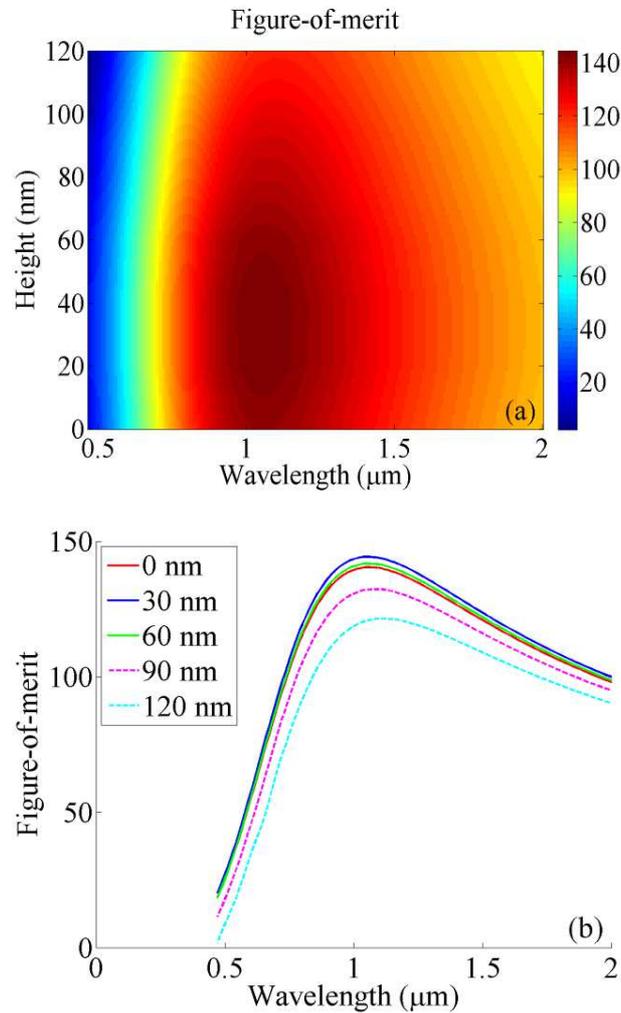

Fig. 8. (Color online) (a) Figure-of-merit versus the wavelength and triangle height. (b) Figure-of-merit versus the wavelength for several triangle heights.

We have calculated the propagation distance and the figure-of-merit of the 120 *nm* wide triangular nanoridge waveguide versus the triangle height at the 1.55 $\mu m$ wavelength. The results are shown in Fig. 9. As the triangle height increases, both the propagation distance and the figure-of-merit increase when the height is smaller than 20 *nm*, and reach a plateau when the height is between 20 - 50 *nm*, then are seen to dramatically decrease. It is concluded that for the 120 *nm* wide silver triangular cross section nanoridge waveguides discussed here, the best performance nanoridge plasmon waveguide is the nanoridge with the triangle height between 20 – 50 *nm*.

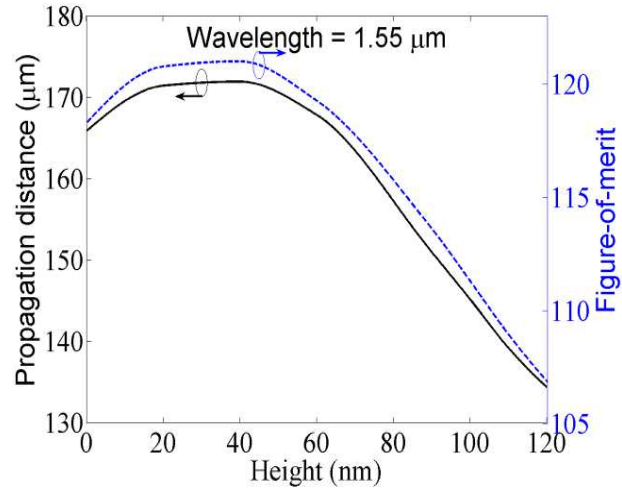

Fig. 9. (Color online) Propagation distance and figure-of-merit of 120 *nm* wide triangular nanoridge plasmon waveguide mode versus the triangle height at 1.55 *μm* wavelength.

We also have calculated the propagation distance and the figure-of-merit of the triangular cross section silver nanoridge waveguide for different triangle height and nanoridge width at the 1.55 *μm* wavelength. The results are shown in Fig. 10. The horizontal axis is the height of the triangular ridge from 0 to 1000 nm. The vertical axis is the width of the nanoridge from 50 to 500 nm. It can be seen clearly that as the height of the triangular ridge increases, the propagation distance and the figure-of-merit increase first, and then decrease. The white dashed lines in Fig. 10 indicate the triangular cross sections in which the three wedge angles are the same at 120 degree. The propagation distance and figure-of-merit reach the maximal values along this line. This is because the plasmon mode energy is more equally distributed among three coupled wedge modes.

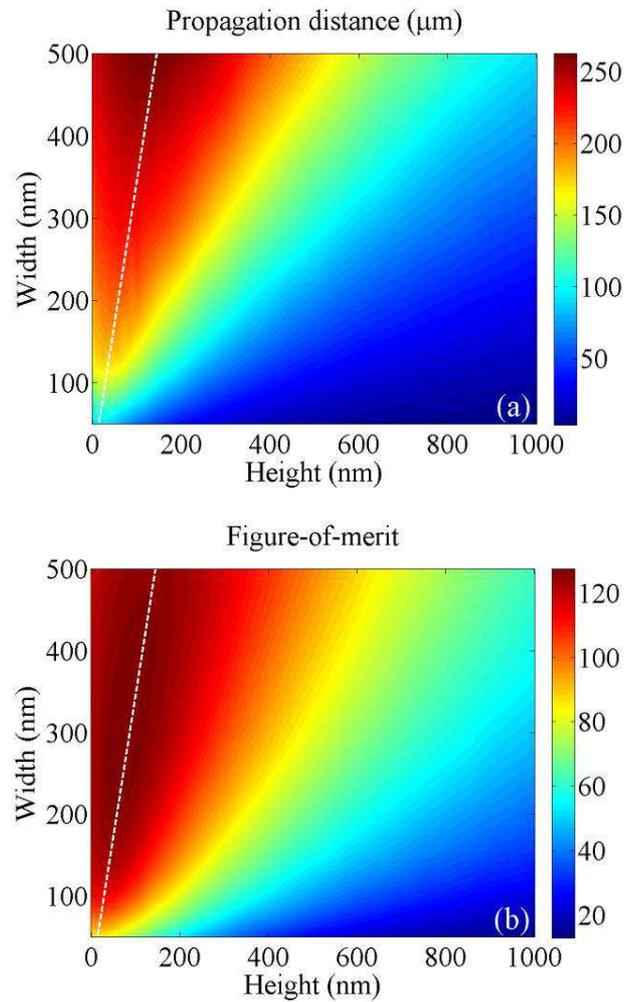

Fig. 10. (Color online) Propagation distance and figure-of-merit of triangular cross section nanoridge waveguide mode versus the triangle height and the width of the nanoridge at 1.55 $\mu m$ wavelength.

## 3. Inverted triangular cross section nanoridge plasmon waveguides

The inverted triangular nanoridge waveguides are systematically investigated in this section, whose 3D view and cross section are plotted in Fig. 11(a) and Fig. 11(b), respectively. An inverted triangle with depth ($d$) in the $y$-direction is removed from the top of the nanoridge plasmonic waveguide.

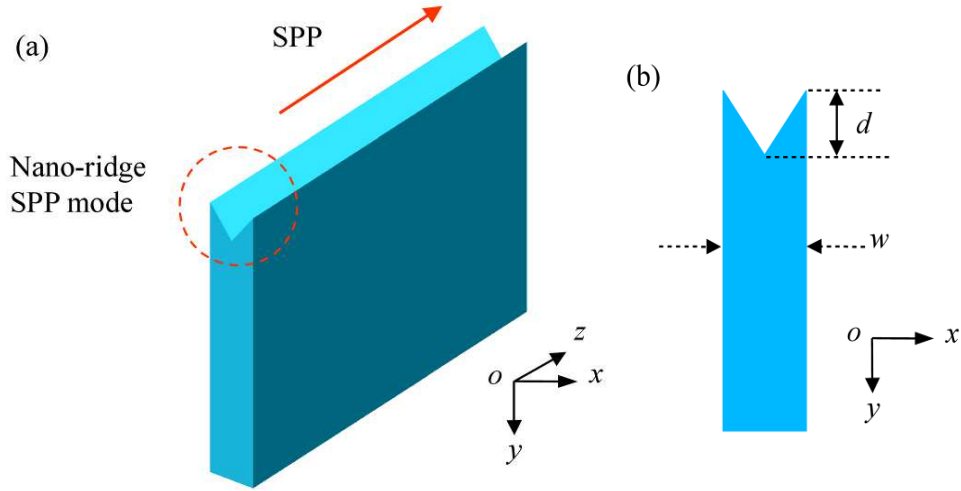

Fig. 11. (Color online) (a) 3D view of the inverted triangular nanoridge plasmon waveguide. (b) Cross section of the inverted triangle nanoridge plasmon waveguide.

We calculated the electric field intensity distributions of the fundamental plasmon propagating modes supported by inverted triangular nanoridges of different inverted triangle depths ($d$ = 0 *nm*, 20 *nm*, 40 *nm*, 60 *nm*, 80 *nm*, and 100 *nm*) at the 1.55 $\mu m$ wavelength. The widths of the nanoridges are all at 120 *nm*. The results are shown in Fig. 12. It can be seen that at any depth of the inverted triangular nanoridge, the mode has only two hot spots. Unlike V-shape trenches in metal surfaces [19-26], there is no hot spot at the bottom of the inverted triangular nanoridge waveguide. The electrons in the inverted triangular nanoridge waveguide are unstable at the bottom of the inverted triangle, so the electrons will move to the two sharp corners on the top of this structure. And it is apparent that as the depth increases, the hot spots of the modes do not move, but the modes become more tightly confined at the two hot spots on the top of the inverted triangular nanoridges. The plasmon mode of the inverted triangular nanoridge can be considered as a hybrid mode consisting of two wedge plasmon modes.

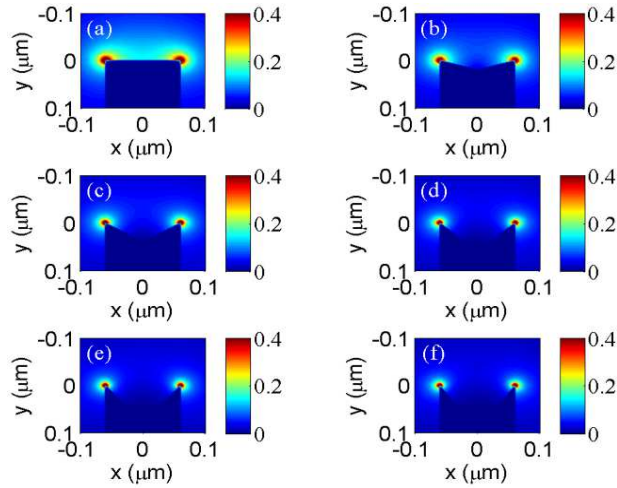

Fig. 12. (Color online) Electric field intensity distributions of the 120 *nm* wide inverted triangular nanoridge plasmon waveguide with the inverted triangle depth equals (a) 0 *nm*, (b) 20 *nm*, (c) 40 *nm*, (d) 60 *nm*, (e) 80 *nm*, and (f) 100 *nm*, at 1.55 $\mu m$ wavelength.

The mode dispersion curves for inverted triangular silver nanoridges of different inverted triangle depths are shown in Fig. 13. The black solid line and dashed line are the dispersion curves of the light line in air, and the silver-air flat plasmon mode, respectively. The inverted triangle depth increases from the zero (the flat-top ridge) to 120 *nm*. As the frequency or the inverted triangle depth increases, the plasmon mode dispersion curves of all ridges drift away from the light line in air, leading to the slower group velocity and tighter mode confinement at the two sharp corners.

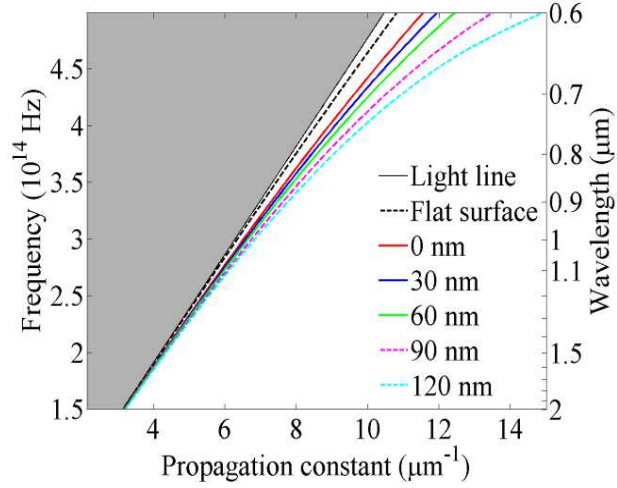

Fig. 13. (Color online) Dispersion curves of the silver inverted triangular nanoridge plasmon waveguides of different inverted triangle depths and the comparison with the dispersion curve of the silver-air flat plasmon mode.

The real and imaginary part of the inverted triangular nanoridge mode index versus the wavelength for several different inverted triangle depths ($d = 0$ *nm*, 30 *nm*, 60 *nm*, 90 *nm*, and 120 *nm*) are plotted in Fig. 14, where the black dashed lines are those of the silver-air flat plasmon mode. It can be seen that as the inverted triangle depth decreases or the wavelength increases, both the real and imaginary part of the inverted triangular nanoridge mode index are reduced, indicating the reduction of the mode confinement and propagation loss.

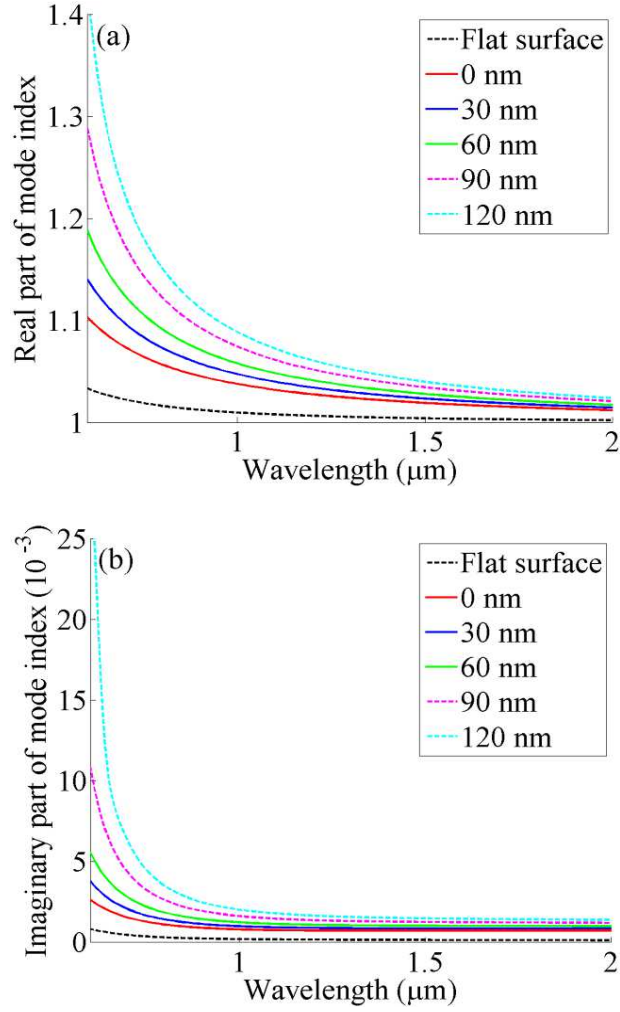

Fig. 14. (Color online) (a) Real part of the mode index versus wavelength for different inverted triangle depths and the comparison with the real part of the silver-air flat plasmon mode index. (b) Imaginary part of the mode index versus wavelength for different inverted triangle depths and the comparison with the imaginary part of the silver-air flat plasmon mode index.

The real part and imaginary part of the mode index versus the inverted triangle depth at the telecommunication wavelength of 1.55 $\mu m$ are shown in Fig. 15. It can be seen that as the inverted triangle depth increase, both of the real and imaginary part of the mode index increase rapidly, which indicates quick rise of attenuation and confinement.

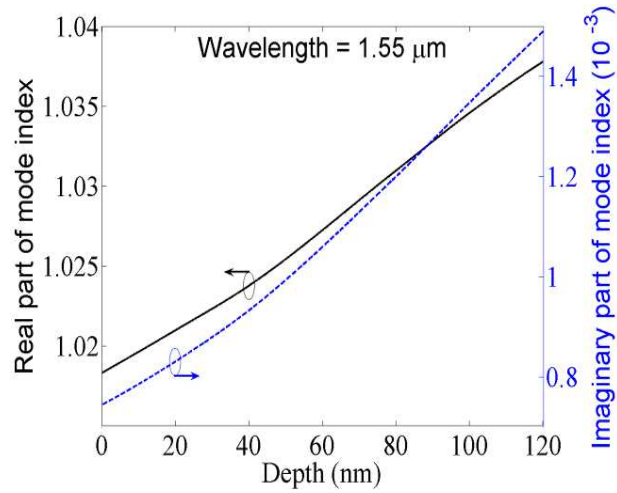

Fig. 15. (Color online) Real and imaginary part of the mode index as a function of the inverted triangle depth at 1.55 $\mu m$ wavelength.

The propagation distances are calculated for nanoridge waveguides of different inverted triangle depths at different free space wavelength. The results are shown in Fig. 16(a-b). It can be seen that as the wavelength increases, the propagation distance increases. The inverted triangular nanoridge can be considered as a hybrid mode consisting of two wedge plasmon modes. As the inverted triangle depth increases, the coupled wedge plasmon modes of the inverted triangular nanoridge have larger loss, so the nanoridge plasmon mode has the smaller propagation distance.

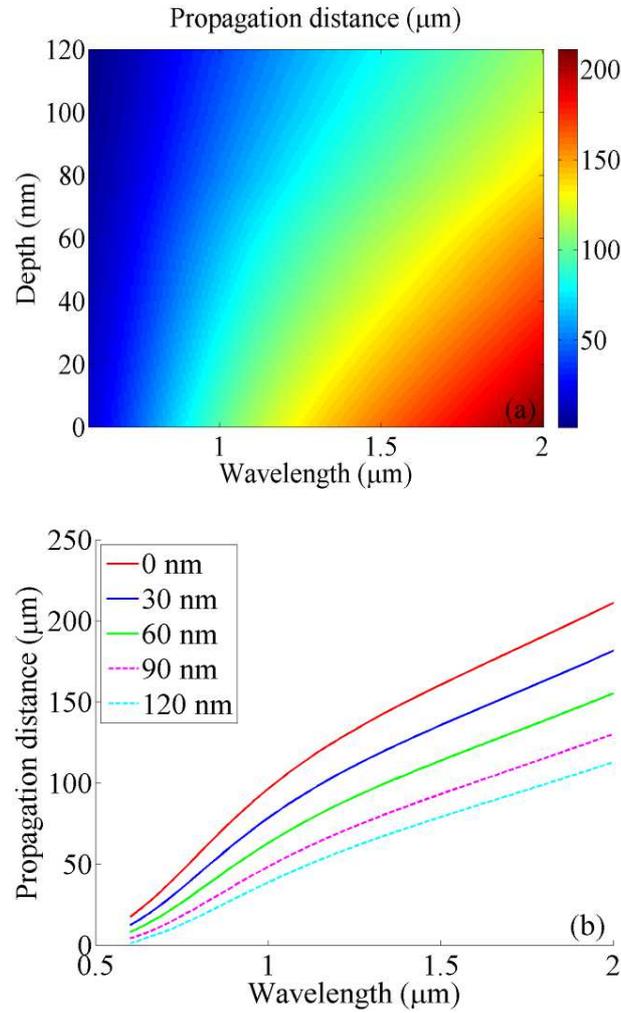

Fig. 16. (Color online) (a) Propagation distance versus wavelength and inverted triangle depth. (b) Propagation distance of the inverted triangular nanoridge mode as a function of free space wavelength for different inverted triangle depths.

The mode size of the inverted triangular nanoridge waveguide versus the free space wavelength and inverted triangle depth is shown in Fig. 17(a), while Fig. 17(b) shows that the mode size versus the free space wavelength for the inverted triangular nanoridge of different inverted triangle depths. Here is seen that as the wavelength increases or the inverted triangle depth decreases, the mode size increases.

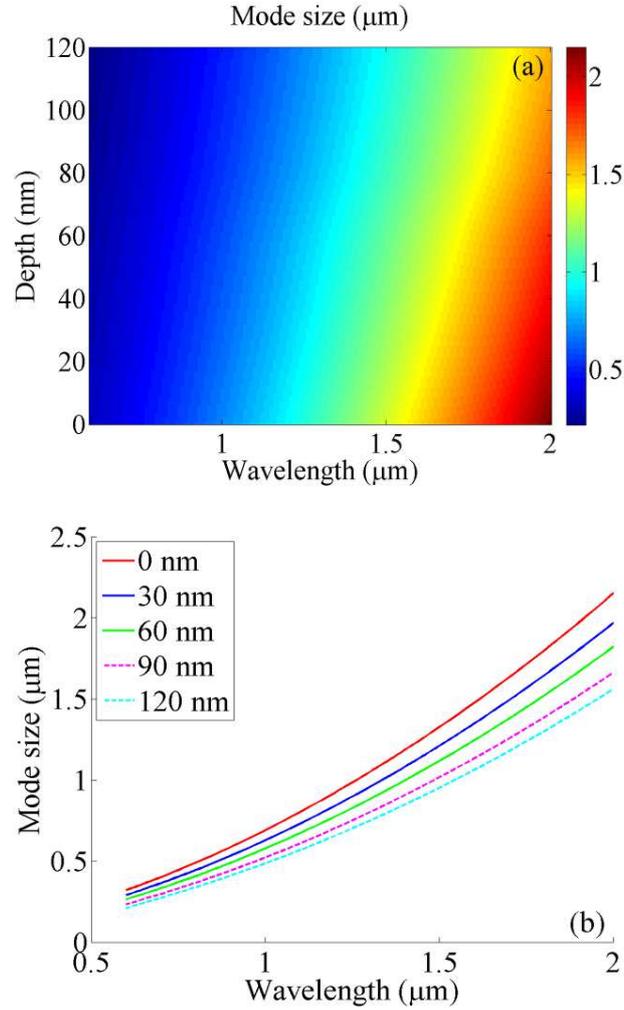

Fig. 17. (Color online) (a) Mode size versus the wavelength and inverted triangle depth. (b) Mode size of the inverted triangular nanoridge plasmon mode as a function of free space wavelength for different inverted triangle depths.

To characterize the trade-off between the attenuation and the confinement of the inverted triangular nanoridge waveguide, we calculate the figure-of-merit as defined in Eq. 2. Fig. 18(a) shows the figure-of-merit of the inverted triangular nanoridge waveguide versus wavelength and inverted triangle depth, while Fig. 18(b) shows the figure-of-merit versus the wavelength for several different inverted triangle depths. It can be seen that as the wavelength increases, the figure-of-merit first increases and then decreases. For all inverted triangular nanoridge of various depths considered in this study, the figure-of merit reaches the maximum at 1.05 $\mu m$ wavelength. It also can be seen that as the inverted triangle depth increases, the figure-of-merit decreases rapidly.

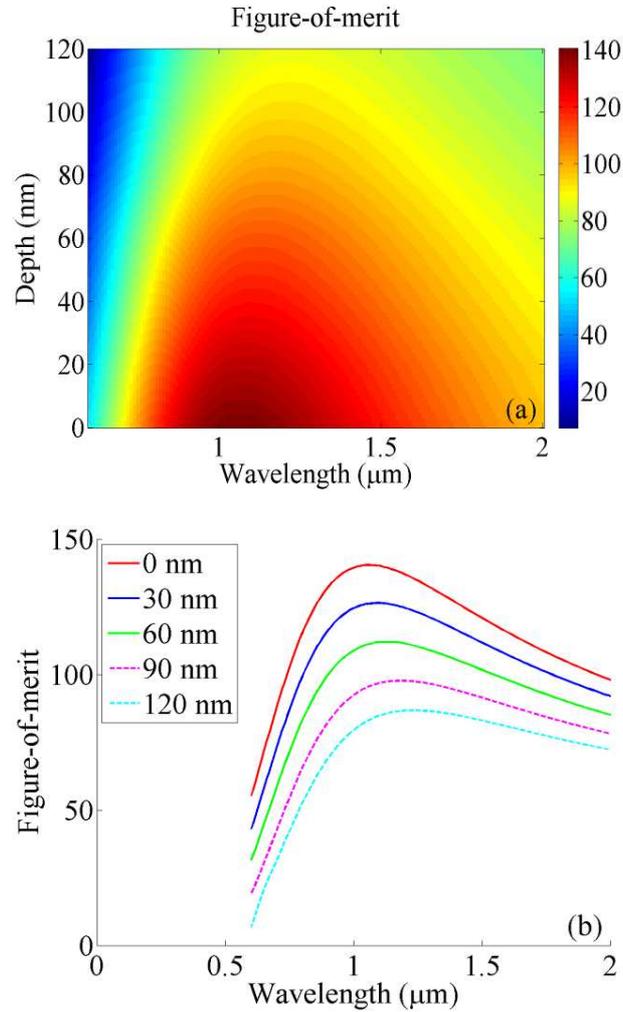

Fig. 18. (Color online) (a) Figure-of-merit versus the wavelength and inverted triangle depth. (b) Figure-of-merit versus the wavelength for several inverted triangle depths.

We also calculate the propagation distance and the figure-of-merit of the inverted triangular nanoridge waveguide versus the inverted triangle depth at the 1.55 $\mu m$ wavelength. The results are shown in Fig. 19. As the inverted triangle depth increases, both of the propagation distance and the figure-of-merit are seen to decrease dramatically. Therefore, when the nanoridge waveguides are fabricated, the concave on the top of the nanoridges should be avoided.

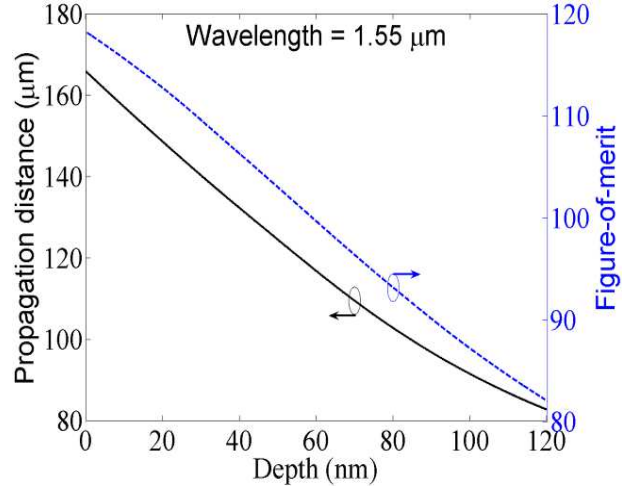

Fig. 19. (Color online) Propagation distance and figure-of-merit of inverted triangular nanoridge plasmon waveguide mode versus the inverted triangle depth at 1.55 *μm* wavelength.

## 4. Summary

We investigated the propagating plasmon modes supported by silver nanoridge waveguides with triangular and inverted triangular cross sections. We calculated the mode dispersions, propagation distances, mode sizes, and the figure-of-merits of these nanoridge plasmon modes. It is found that as the height of the triangle increases, the propagation distance and the figure-of-merit first increase, and after reaching a plateau decrease quickly. When the triangle height falls between 20 – 50 *nm* for 120 nm wide silver ridges, the ridge plasmon modes have the longest propagation distance and the highest figure-of-merit. The triangular ridge waveguides with 20 – 50 *nm* triangle height have better performance in terms of the propagation distance and figure-merit than flat-top nanoridge waveguides previously investigated [53]. By varying the nanoridge width and the triangle height, the propagation distance and the figure-of-merit reach maximal values when the three wedge angles approach the same at 120 degree. When the three wedge angles are at 120 degree, the plasmon mode energy of triangular cross section ridge waveguide is more equally distributed among three equal angle wedges. For the inverted triangular nanoridge waveguides, it is found that the propagation distance and the figure-of-merit decrease dramatically when the inverted triangle depth increases. This indicates that inverted triangular nanoridge plasmon waveguides have poorer performance when compared to the flat-top nanoridge plasmon waveguides. This work suggests, that for building lower loss and tighter confinement nanoridge plasmon waveguide circuits, a triangular nanoridge waveguide with the top wedge angle of 120 degree, represents an optimal trade-off between the mode confinement and propagation distance.

## Acknowledgment

Junpeng Guo acknowledges the support from the ASEE-Air Force Office of Scientific Research (AFOSR) Summer Faculty Fellowship Program. This work was also partially supported by the National Science Foundation (NSF) through the award NSF-0814103 and the National Aeronautics and Space Administration (NASA) through the grant NNX07 AL52A.